\documentclass[]{article}

\usepackage{arxiv}

\usepackage{amsmath}
\usepackage{amssymb}

\usepackage[utf8]{inputenc} 
\usepackage[T1]{fontenc}    
\usepackage{hyperref}       
\usepackage{url}            
\usepackage{booktabs}       
\usepackage{amsfonts}       
\usepackage{nicefrac}       
\usepackage{microtype}      
\usepackage{cleveref}       
\usepackage{lipsum}         
\usepackage{graphicx}
\usepackage[square,numbers,super,sort&compress]{natbib}
\usepackage{doi}

\usepackage{makecell}
\usepackage{arydshln}

\usepackage{tikz}                    
\usetikzlibrary{shapes, arrows, positioning, fit, backgrounds, calc, matrix}
\usepackage{caption}                 
\usepackage{float}                   
\usepackage{geometry}                
\usepackage{adjustbox} 
\usetikzlibrary{arrows.meta, shapes.geometric, decorations.pathreplacing}  

\usepackage{rotating}  
\usepackage{pdflscape} 
  
\usepackage{booktabs, threeparttable} 
\usepackage{longtable}
\usepackage{tabularx}
\usepackage[justification=centerlast]{caption} 
\usepackage{subcaption} 
\usepackage{xcolor}

\usepackage{relsize}  
 
\usepackage{subcaption} 

\usepackage{siunitx}   

\usepackage{listings} 

\definecolor{codegreen}{rgb}{0,0.6,0}
\definecolor{codegray}{rgb}{0.5,0.5,0.5}
\definecolor{codepurple}{rgb}{0.58,0,0.82}
\definecolor{backcolour}{rgb}{0.95,0.95,0.92}

\lstdefinestyle{promptstyle}{
	backgroundcolor=\color{backcolour},   
	commentstyle=\color{codegreen},
	keywordstyle=\color{magenta},
	numberstyle=\tiny\color{codegray},
	stringstyle=\color{codepurple},
	basicstyle=\ttfamily\footnotesize,
	breakatwhitespace=false,         
	breaklines=true,                 
	captionpos=b,                    
	keepspaces=true,                 
	numbers=left,                    
	numbersep=5pt,                  
	showspaces=false,                
	showstringspaces=false,
	showtabs=false,                  
	tabsize=2,
	frame=single,
	rulecolor=\color{black}
} 

\lstdefinestyle{pythonstyle}{
	backgroundcolor=\color{backcolour},   
	commentstyle=\color{codegreen},
	keywordstyle=\color{magenta},
	numberstyle=\tiny\color{codegray},
	stringstyle=\color{codepurple},
	basicstyle=\ttfamily\footnotesize,
	breakatwhitespace=false,         
	breaklines=true,                 
	captionpos=b,                    
	keepspaces=true,                 
	numbers=left,                    
	numbersep=5pt,                  
	showspaces=false,                
	showstringspaces=false,
	showtabs=false,                  
	tabsize=2,
	frame=single,
	rulecolor=\color{black},
	language=Python
}

\newcommand{\minitrend}[3]{%
	\begin{tikzpicture}[baseline={(0,0)}, x=0.4cm, y=0.2cm]
		\pgfmathsetmacro{\maxval}{max(#1,#2,#3)+0.0001}
		\pgfmathsetmacro{\minval}{min(#1,#2,#3)}
		\pgfmathsetmacro{\range}{\maxval-\minval}
		\pgfmathsetmacro{\yone}{(#1-\minval)/\range}
		\pgfmathsetmacro{\ytwo}{(#2-\minval)/\range}
		\pgfmathsetmacro{\ythree}{(#3-\minval)/\range}
		
		\draw[gray!45, very thin, densely dotted] (0,0) -- (2,0);
		\draw[violet!70!black, semithick] (0,\yone) -- (1,\ytwo) -- (2,\ythree); 
		plot[smooth, tension=0.7] coordinates {(0,\yone) (1,\ytwo) (2,\ythree)};
		\fill[violet!70!black] (0,\yone) circle (1.2pt);
		\fill[violet!70!black] (1,\ytwo) circle (1.2pt);
		\fill[violet!70!black] (2,\ythree) circle (1.2pt);
	\end{tikzpicture}%
}

\crefname{section}{Sec.}{Secs.}
\Crefname{section}{Section}{Sections}
\Crefname{table}{Table}{Tables}
\crefname{table}{Tab.}{Tabs.}

\newcommand{\dpaperbib}{pop-enhanced-latent-code}

\newcommand*{\zsubsection}{\subsection} 
 
\newcommand{\TheTitle}{In-batch Relational Features Enhance Precision in An Unsupervised Medical Anomaly Detection Task}
\newcommand{\TheTitleShort}{In-batch Relational Features Enhance Precision in An Unsupervised Medical Anomaly Detection Task}

\title{\TheTitle}

\date{}
 
\usepackage{authblk}

\newbox{\orcid}\sbox{\orcid}{\includegraphics[scale=0.06]{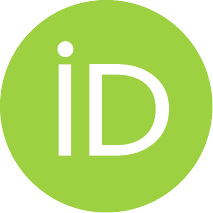}} 
\author[1]{%
	\href{https://orcid.org/0009-0000-2080-4979}{\usebox{\orcid}\hspace{1mm}P.~Bilha Githinji}%
}  
\author[2]{%
\href{}{\usebox{\orcid}\hspace{1mm}Ijaz Gul}%
}   
\author[2]{%
\href{}{\usebox{\orcid}\hspace{1mm}Lian Zhang}%
}   
\author[1]{%
\href{}{\usebox{\orcid}\hspace{1mm}Jinhao Xu}%
}   
\author[1]{%
	\href{https://orcid.org/0000-0001-7336-7848}{\usebox{\orcid}\hspace{1mm}Peiwu Qin}%
} 
\author[1]{%
\href{}{\usebox{\orcid}\hspace{1mm}Dongmei Yu}%
}   
\affil[1]{Guangdong Provincial Laboratory of Traditional Chinese Medicine, Guangdong, China}
\affil[2]{The First Hospital of Hebei Medical University, Shijiazhuang, China}

\renewcommand{\headeright}{} 
\renewcommand{\undertitle}{} 
\renewcommand{\shorttitle}{\TheTitleShort}

\hypersetup{
pdftitle={\TheTitle},
pdfsubject={},
pdfauthor={P. Bilha.~Githinji, Et al.},
pdfkeywords={Anomaly Detection, Relational Regularization, Latent Space Regularization, Graph Learning, Population-aware Embedding, False Positive Reduction},
}

\begin{document}
\maketitle

\begin{abstract}	

Confounding pathology with normal anatomical variation remains a significant challenge in unsupervised medical-image anomaly detection, resulting in numerous false positives.  
To enhance integration of healthy variation, we augment the latent representation of a CNN autoencoder with contextual similarities within a normal cohort through batch-wise hypergraph estimation and a shared-weights graph convolution layer, producing a population-aware embedding. 
On a heterogeneous brain-tumor dataset of 2D MRI scans, the method improves separability between healthy and pathological samples, achieving an AUC-ROC of 0.90 (95\% CI 0.84-0.95, 5.7\% absolute gain), and a 16\% absolute improvement in average precision (0.78 AP, 95\% CI 0.66-0.89), thereby lowering false-positive rates. 
Moreover, both anomaly detection and downstream tumor versus no-tumor classification performance improve with the size of the mini-batch context captured in the augmented representation, suggesting a tunable lever for integrating healthy variation.

\end{abstract}

\section{Introduction}
\label{sec:dpintro}

 Unsupervised anomaly detection enables the identification of lesions and rare pathologies without the burden of extensive annotation.
 The utility of reconstruction-based unsupervised anomaly detection hinges on balancing the expressiveness required to model complex healthy tissue against the restrictiveness needed to fail on anomalies~\cite{baur_autoencoders_2021,tschuchnig_anomaly_2022,kim_anomaly_2026}.
 High-capacity models will generalize too well and accurately reconstruct anomalies, despite being trained exclusively on healthy data. Conversely, overly constraining the latent space results in the loss of fine-grained anatomical details, conflating benign variance with pathology and inflating false positive rates.

 To disentangle acceptable biological variance from pathological deviation, we propose a method that leverages the relational structure of a healthy cohort to regularize the latent representation of an unsupervised anomaly detection image auto-encoder. 
 Unlike standard training, where image samples are encoded independently~\cite{baur_autoencoders_2021,hassanaly_unsupervised_2024}, our approach dynamically constructs a similarity-based hypergraph within each training mini-batch to define the manifold of plausible healthy variation. 
 We utilize a graph convolutional neural network (GCN) layer within the bottleneck of a convolutional neural network (CNN) auto-encoder (AE) to aggregate relational features from neighboring healthy samples, augmenting the latent code with population-aware constraints. 
 By requiring that the latent codes be coherent with the healthy cohort, our method ensures that the model is expressive enough to reconstruct complex healthy anatomy yet restrictive enough to fail on anomalies that lack support in the training cohort, thereby improving detection separability and reducing false positive rates.

 Our specific contributions include
 
 \begin{itemize}  
 	\item{\textit{Population-aware feature refinement}.
 		We propose a module that transforms independent sample-based latent codes into population-aware representations. By dynamically estimating similarity-based hypergraphs within training mini-batches and utilizing a shared-weight GCN layer, we aggregate relational features from $k$-nearest neighbors and project an augmented latent code to the decoding unit. This integration is architecturally non-intrusive, preserving the base auto-encoder structure. 
 	}
 	
 	\item{\textit{Mitigation of false positives}. 
 		We empirically demonstrate the feasibility of a batch-wise implementation and that the method particularly mitigates false positive rates. The proposed method delivers a 16.0\% absolute improvement in average precision (AP, 25.9\% relative improvement) and a 5.7\% absolute gain in AUC-ROC (6.7\% relative gain). Mean AP = $0.78$ [95\% CI: $0.66 - 0.89$], mean AUC = $0.90$ [95\% CI: $0.84 - 0.95$].  
 	}
 	
 	\item{\textit{Manifold structure analysis}. 
 		We systematically assess the impact of the neighborhood context size $k$. We find that a broader population-aware augmentation attains a more discriminative latent space structure and that there is a monotonic improvement in performance metrics as the context size increases. Moreover, an insufficient context does not degrade the host AE.
 	} 
 \end{itemize}


\section{Prior Works}
\label{sec:dbackground}

Prior approaches to improving the performance of unsupervised anomaly detection entail constraining the capacity of the model through architectural asymmetry, creating restrictive bottlenecks, and enhancing the latent representation with prototypical features of normalcy. 
Architectural asymmetry intentionally cripples the decoder's capacity to reconstruct samples (e.g., removing skip connections) or employs an architecturally different decoder (e.g., transformers) with an inductive bias that differs from that of the encoder, in effect degrading reconstruction of anomalous regions~\cite{lu_anomaly_2024,hassanaly_unsupervised_2024,pinaya_unsupervised_2022,xing_memory-augmented_2025}.
Enhancing latent representations with prototypical healthy features has been achieved through integration of discrete memory banks or vector quantization~\cite{gong_memorizing_2019,zhou_proxy-bridged_2021,xing_memory-augmented_2025}. During training, these methods learn a discrete set of latent features that store snapshots of prototypical normal patterns. While effective, these methods are constrained by the size of the memory bank, which may additionally hinder the ability to capture continuous variations.

Our work focuses on enhancing the latent representation of an unsupervised anomaly detection AE by integrating shared features of normalcy using graph convolution within a training batch.
Unlike memory modules that represent the population norm with a static set of prototypes, our approach learns to dynamically infer relational sub-structures of healthy cohorts and learn the continuous rules of normal variation within the shared weights of the GCN. 
Furthermore, while graph neural networks (GNNs) and hypergraphs have successfully modeled population relationships for disease classification~\cite{ijcai2024p903,linguraru_customized_2024,linguraru_inter-intra_2024}, this work explores their application as a dynamic regularization mechanism for pixel-wise reconstruction in unsupervised anomaly detection. 
Our approach structures the latent space of the AE to better model the manifold of normal variation without auxiliary losses or architectural changes to the host AE.

\section{Materials and Methods}
\label{sec:dpmethod}
   
 \zsubsection{Dataset}
 
 The input dataset is a heterogeneous public brain tumor dataset of 7,023 2-D MRI images collated from three other sources~\cite{noauthor_brain_nodate}. It captures variations in intensity values and covers four disease classes, namely, glioma, meningioma, pituitary tumors, and no-tumor. The no-tumor or \emph{normal} class contains images from the Br35H dataset. We utilize the axial plane images and normalize the intensity range of images to $[0,1]$. Training is conducted with the \emph{normal} samples only, which are additionally augmented with random operations of blurring and masking of several small regions with the median intensity value of the input image.

 \zsubsection{Model architecture}
 The core of the system is a reasonably capable autoencoder (AE) into which we integrate our proposed method and explore its contribution.
 We adopt the ResNet101 backbone and under-specify the decoder function by dropping skip connections from the first layer of the encoder. This serves to regularize against learning the identity function.  
 This AE is also a baseline model.

  \paragraph{Population-aware latent representation.} 
 Our proposed module operates in the latent space of a convolutional auto-encoder. It takes as input the latent code $z_{e} \in \mathbb{R}^{c \times h \times w}$ from the AE encoder, and outputs an augmented but similarly sized latent code $z_{g} \in \mathbb{R}^{c \times h \times w}$ that is then fed to the decoder. 
 For a mini-batch of size $B$, we estimate a k-uniform similarity-based hypergraph and then utilize message passing convolution operation of a single-layer graph convolutional network (GCN) to aggregate the neighborhood features and update the latent code of the sample.
 Using a hyperparameter $k_{\mathrm{perc}} \in [0,1]$, the number of k-nearest neighbors is determined as $k = \lceil B * k_{\mathrm{perc}} \rceil $. 
 The hypergraph $G=\{V,E,W\}$ is then estimated such that samples in a batch constitute the node set $V = \{{v}_i\}_{i=1}^B$, the set of edges $E = \big\{(i, j) \mid j \in \mathcal{N}_k(i)\big\}$ depicts the association between a node $v_{i}$ and its $k$ nearest neighbors $\mathcal{N}_k(i)$, and these edges are uniformly weighted ($W = \mathbf{I}$ is the edges' weight matrix).

 For a sample $v_{i}$, the augmented representation projected back by the proposed module is determined as 
 \[ 
 \mathbf{z}_{g} = \mathbf{\Theta}_{p} \Big(   \mathbf{z}_{e} || \mathbf{\sigma} \Big( \mathbf{z}_{h}\Big) \Big)
 \]
 where $\mathbf{\Theta}_{p}$ is the projection matrix, $||$ is the concatenation operation, and $\mathbf{\sigma}$ is an activation function applied to the output of the message-passing operation during graph convolution. 
 The message-passing operation aggregates features from the sample $v_{i}$ and its neighbors $\mathcal{N}_k(i)$ using the incidence matrix $H$ of the estimated mini-batch hypergraph $G$, generating a neighborhood feature 
 \[ 
 \mathbf{z}_{h} = \mathbf{z}_{e} \mathbf{\Theta}_{0} + D_{v}^{-\frac{1}{2}} H W D_{e}^{-1} H^T D_{v}^{-\frac{1}{2}} \mathbf{Z}^{\mathcal{N}_k(i)}_{e} \mathbf{\Theta}_{1}
 \]
 where $D_v, D_e$ are the degree matrices for the nodes and edges in the hypergraph $G$, respectively, and  $\mathbf{\Theta}_{0}$, and $\mathbf{\Theta}_{1}$ are learnable weight matrices shared across all nodes.

  \begin{figure}[htbp]  
 	\centering  
 	\begin{subfigure}[b]{0.9\textwidth}
 		\centering 
 		\includegraphics[width=\linewidth]{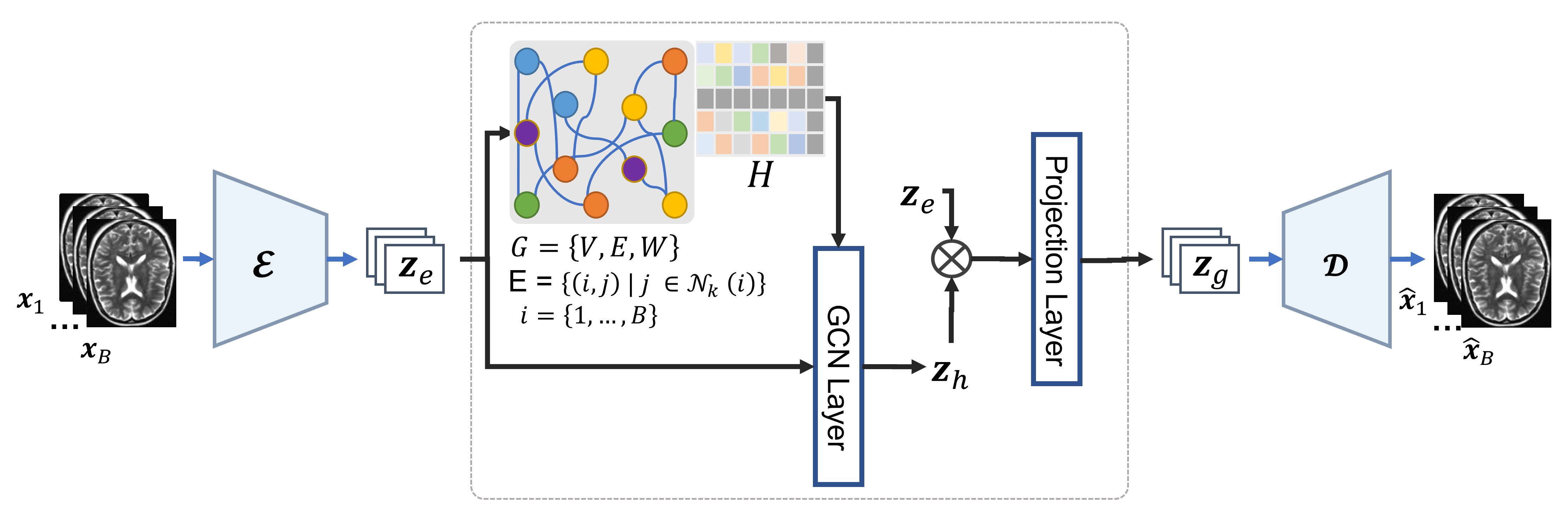} 
 	\end{subfigure} 
 	\caption{Proposed augmentation} 
 	
 	\label{fig:schematic} 
 	
 \end{figure}

 \paragraph{Loss function.}
During training, the module learns the weights $\mathbf{\Theta}_{h}=\{\mathbf{\Theta}_{0}, \mathbf{\Theta}_{1} \}$ that capture relevant population-level features for the task. 
We retain the standard reconstruction loss for medical images, which combines mean squared error (MSE) and structural similarity index metric (SSIM) as shown in ~\Cref{eq:loss}, where $x$ is an input image and $\hat{x}$ is the output image reconstructed by the AE model.
Using a small subset of the training dataset, we empirically determine the values $\lambda_{1} = 1000$ and $\lambda_{2} = 1$ as suitable hyperparameters with high-fidelity reconstructions of the input images.  

	\begin{equation}
		\mathcal{L}_{\text{rec}} (x, \hat{x}) \;=\; \lambda_{1} \,\mathcal{L}_{\text{MSE}} \;+\; \lambda_{2}\,\bigl(1-\text{SSIM}\bigr)
		\label{eq:loss}
	\end{equation}

\paragraph{Training procedure.}
We train on an NVIDIA A100 GPU with 80GB RAM since RAM size is a critical limiting factor for our in-batch operation. Other training hyperparameters are standard and include Adam optimizer with a learning rate of $1\times10^{-4}$, and weight decay of $0.5\times10^{-5}$.

\zsubsection{Evaluation}
We comprehensively assess the contribution of the proposed module, taking into consideration anomaly discriminability performance, quality of the resulting latent space, and downstream utility. 
All analyses are performed on a held-out \emph{normal} evaluation set and \emph{pathological} samples.  

For anomaly discriminability, we evaluate the separability between anomaly scores from healthy and pathological samples using a two-sample Kolmogorov-Smirnov (K-S) test, quantifying the statistical divergence between the score distributions of the two groups. Additionally, we compute standard detection metrics, area under the receiver operating characteristic curve (AUROC or AUC-ROC), and area under the precision-recall curve (AUPRC or AUC-PR), capturing the capacity to reduce false positive rates while maintaining sensitivity.

To assess the structural impact of the proposed module on the learned representations, we compare the augmented latent codes $\mathbf{z}_{g}$ and the standard encoder representation $\mathbf{z}_{e}$. We evaluate clustering performance using standardized metrics such as Silhouette score and Calinski-Harabasz index, and visualize the latent structure with dimensionality reduction techniques (UMAP with PCA). Furthermore, we perform a sensitivity analysis on the hyperparameter $k$, which controls the neighborhood size in the relational hypergraph. 
Moreover, we evaluate the generalizability and discriminative capacity of the latent codes and anomaly scores in direct downstream binary classification of healthy versus pathological samples. We use a simple linear classifier (with 5-fold cross validation) on the latent representations $\mathbf{z}_{e}$ and $\mathbf{z}_{g}$, and employ Youden's $J \;=\; \max_{t} \bigl(\text{TPR}(t) + \text{TNR}(t) - 1\bigr)$ statistic as a decision threshold on the anomaly scores (TPR is true positive rate, while TNR is true negative rate).  

We perform 3,000 independent resampling runs, each drawing 150 samples stratified by class (healthy versus pathological), bootstrapped evaluations when estimating mean values (95\% confidence intervals). Comparative analyses between model configurations employs independent two-tailed Student's $t$-test. Results are considered statistically significant at $\alpha = 0.05$ level ($p$-values $\le 0.05$).

\section{Results}
\label{sec:dresults}

We first verify that our evaluation set provides an unbiased representation of the healthy anatomy. A KS test confirms that the intensity distributions of normal samples in the training and evaluation sets are statistically indistinguishable ($p$-value $= 0.54$). This is critical for brain imaging anomaly detection, where discriminative signals are predominantly intensity-based.

\begin{figure}[htbp]  
    \centering  
    \begin{subfigure}[b]{0.49\textwidth}
        \centering 
        \includegraphics[width=\linewidth]{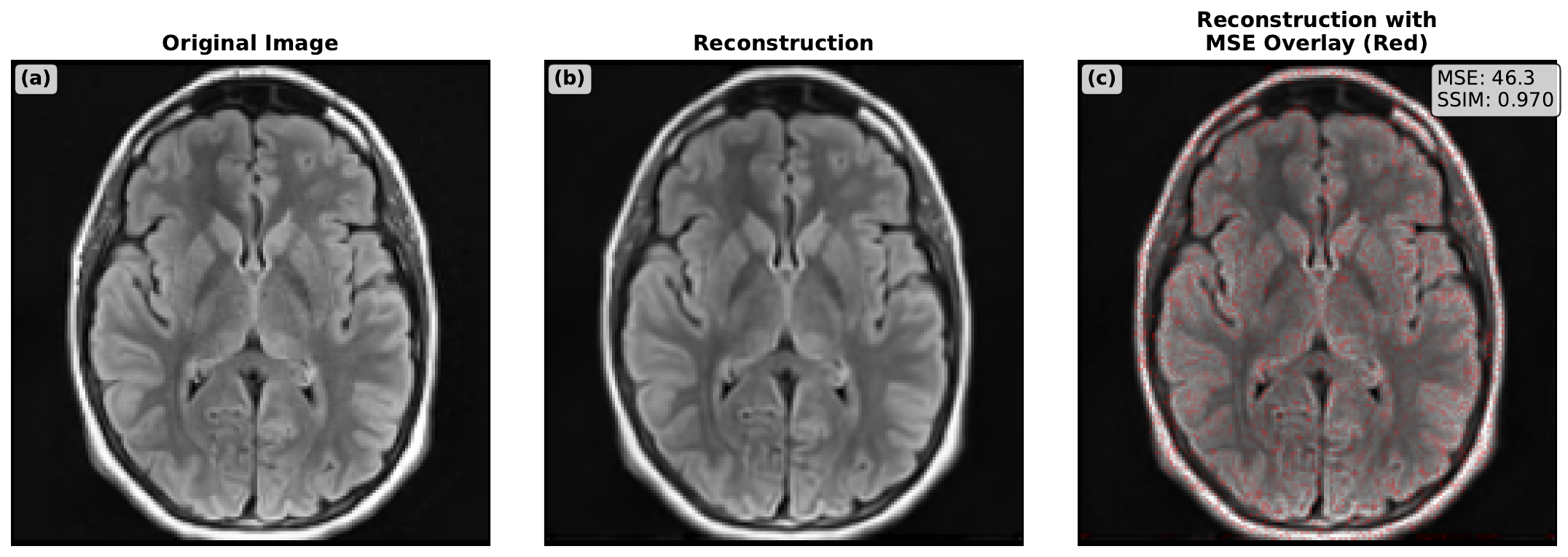}
\end{subfigure}
\begin{subfigure}[b]{0.49\textwidth}
        \centering 
        \includegraphics[width=\linewidth]{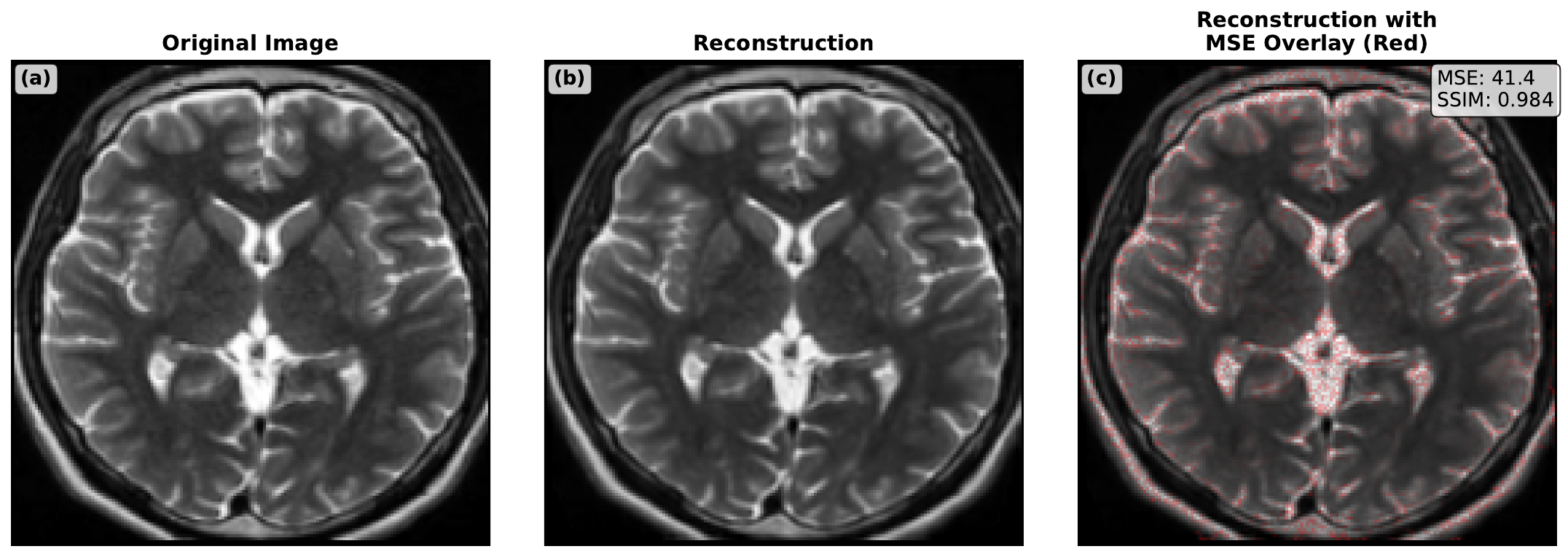}
\end{subfigure}
\hfill
\begin{subfigure}[b]{0.49\textwidth}
        \centering 
        \includegraphics[width=\linewidth]{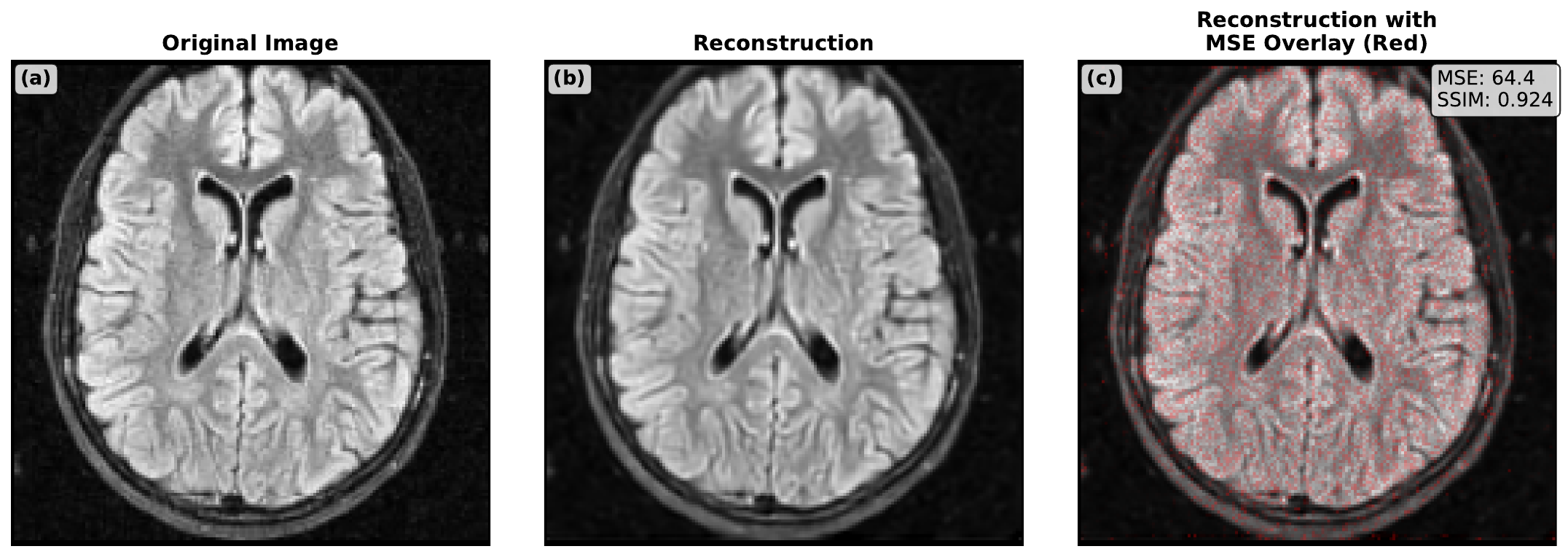}
\end{subfigure}
\begin{subfigure}[b]{0.49\textwidth}
        \centering 
        \includegraphics[width=\linewidth]{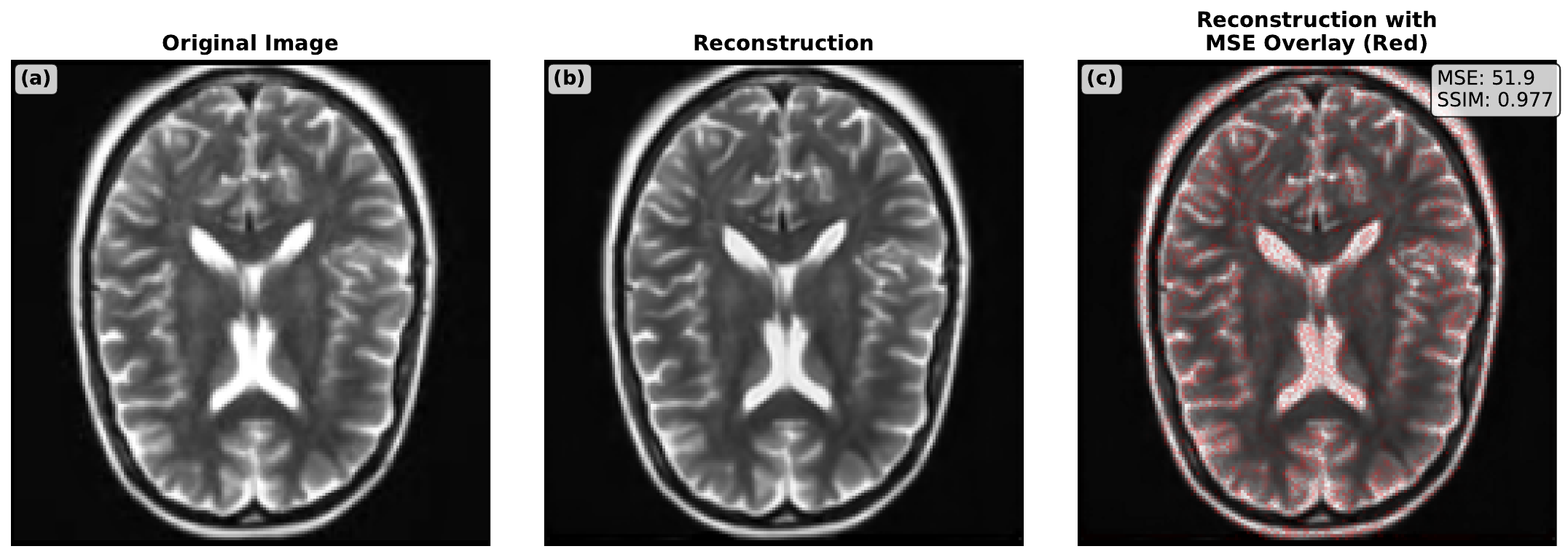}
\end{subfigure}
\hfill

    \caption{Plausible reconstructions} 
     
    \label{fig:ae-plausible} 
      
    \end{figure}

Additionally, we demonstrate that the baseline auto-encoder produces plausible reconstructions and is competitive.
~\Cref{fig:ae-plausible} shows sample input images alongside their reconstructions. We visualize reconstruction errors as overlays of the pixel-wise residual map (squared error, $(x - \hat{x})^{2}$ ) in red. 
Furthermore,~\Cref{tbl:sota-perf} compiles published results from relevant brain anomaly detection studies. The field has predominantly reported area under the receiver operating curve (AUROC) results, which address overall separability performance. We observe that our baseline autoencoder achieves AUROC performance comparable to prior works, thus offering a reasonable control for the proposed method.

\begin{table}[htbp!]
\centering
\caption{Reported baseline and memory-based SOTA performance (specific to Brain MRI)}
\label{tbl:sota-perf}
\small
\begin{tabular}{p{0.6in}p{0.6in}p{0.6in}p{0.6in}p{0.6in}p{0.6in}p{0.6in}p{0.6in}p{0.6in}}
\toprule
  & \small{Standard Auto-Encoder (AE)} & \small{f-AnoGAN} & \small{MemAE} & \small{Proxy-Bridge} & \small{MP-DRA} & \small{Tri-VAE} & \small{Our \newline Baseline AE} & \small{Our \newline Module} \\
\midrule
AUROC & 0.77 & 0.82 & 0.85 & 0.85 & 0.90 & 0.97 & 0.84 & 0.90 \\
AUPRC & 0.68 & - & - & - & - & 0.46 & 0.62 & 0.78 \\
\textit{sources} & \cite{baur_2019,zhou_proxy-bridged_2021} & \cite{schlegl_f-anogan_2019,zhou_proxy-bridged_2021} & \cite{zhou_proxy-bridged_2021,gong_memorizing_2019} & \cite{zhou_proxy-bridged_2021} & \cite{shao_mp-dra_2025} & \cite{wijanarko_tri-vae_2024} &  &  \\
\bottomrule
\end{tabular}
\end{table}


 \zsubsection{Separability of healthy and pathological samples}
 
\begin{figure}[htbp]  
    \centering  
    \begin{subfigure}[b]{0.32\textwidth}
        \centering 
        \includegraphics[width=\linewidth]{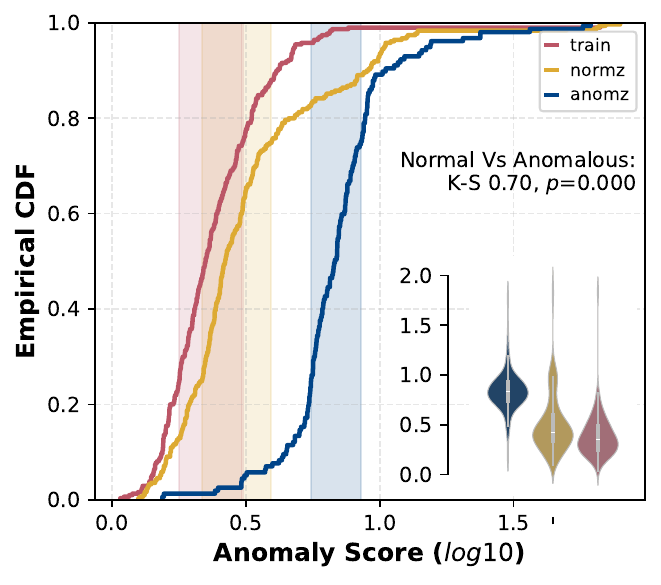}
\subcaption{Plain AE ($k=0$)}\label{fig:ad-dist-vanilla}\end{subfigure}
\begin{subfigure}[b]{0.32\textwidth}
        \centering 
        \includegraphics[width=\linewidth]{04-keffect--distz-obspop\_35-ecdf-vio}
\subcaption{AE+Small-$k$}\label{fig:ad-dist-obspop-35}\end{subfigure}
\begin{subfigure}[b]{0.32\textwidth}
        \centering 
        \includegraphics[width=\linewidth]{04-keffect--distz-obspop\_70-ecdf-vio}
\subcaption{AE+Large-$k$}\label{fig:ad-dist-obspop-70}\end{subfigure}
\hfill

    \caption{Distribution of anomaly scores} 
     
    \label{fig:ad-scores-dist} 
      
    \end{figure}

 Here, we examine the distributions of the anomaly scores, comparing the baseline plain AE to the AE when configured with our population-aware context module. 
 We consider both how the healthy evaluation samples and the pathological samples compare to each other and to the training anomaly scores. This is visualized by distribution plots in~\Cref{fig:ad-scores-dist}. These plots are additionally annotated with the K-S test results for healthy versus pathological distributions. 
 
 For both the baseline AE and AE with our proposed module, the distribution of the healthy evaluation samples aligns closely with that of the training set.  
 Moreover, we see that the distance between the healthy and pathological distributions is greater when the AE is augmented with the proposed solution, attaining a statistically significant KS statistic of $0.76$ compared to $0.70$ under plain AE. Incorporating the population-aware context increases the statistical separation between healthy and pathological distributions, suggesting enhanced capacity to distinguish pathological deviations.

\begin{figure}[htbp]  
    \centering  
    \begin{subfigure}[b]{0.24\textwidth}
        \centering 
        \includegraphics[width=\linewidth]{04-keffect--auc-curves\_\_loss-train\_valid\_norm\_anom\_group-roc-base}
\subcaption{AUC-ROC for anomaly scores}\label{fig:auc-roc}\end{subfigure}
\begin{subfigure}[b]{0.24\textwidth}
        \centering 
        \includegraphics[width=\linewidth]{04-keffect--auc-curves\_\_loss-train\_valid\_norm\_anom\_group-pr-base}
\subcaption{AUC- PR for anomaly scores}\label{fig:auc-pr}\end{subfigure}
\begin{subfigure}[b]{0.24\textwidth}
        \centering 
        \includegraphics[width=\linewidth]{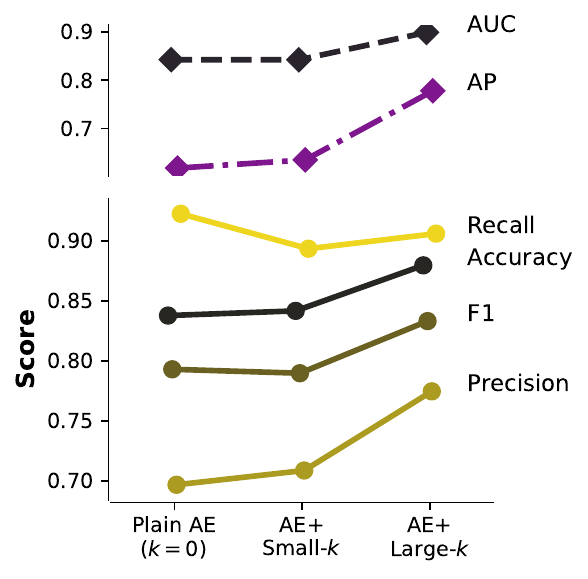}
\subcaption{Thresholding classifier (on anomaly score)}\label{fig:dstream-acc-thresh}\end{subfigure}
\begin{subfigure}[b]{0.24\textwidth}
        \centering 
        \includegraphics[width=\linewidth]{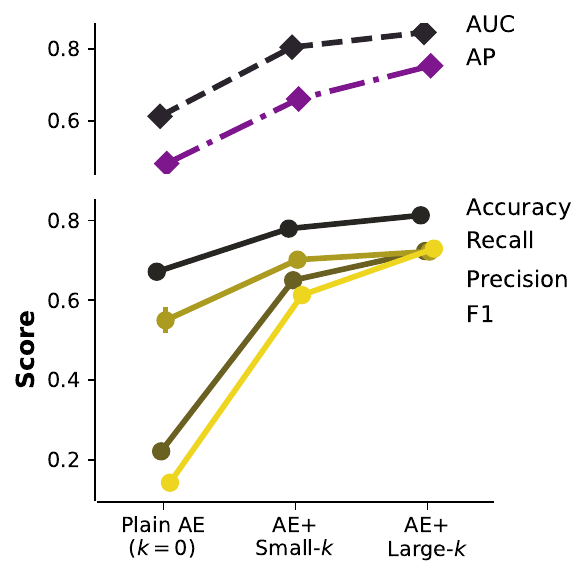}
\subcaption{Logistic regression classifier (on latent code)}\label{fig:dstream-acc-logit}\end{subfigure}
\hfill

    \caption{Anomaly detection (a and b) and downstream classification (c and d) performance} 
     
    \label{fig:ad-roc-pr-acc-all} 
      
    \end{figure}

\zsubsection{Precision performance}
  We additionally evaluate the discriminative capacity of the model configurations using both AUC-ROC and average precision (AP). As shown in~\Cref{fig:ad-roc-pr-acc-all} and in~\Cref{tbl:acc-perf-ci}, where confidence intervals and statistical tests are detailed, the augmented AE significantly outperforms the baseline. 
  The plain AE achieves an AUC score of 0.84 (95\% CI: 0.77 - 0.91), while augmenting the latent representation of the AE improves AUC to 0.90 (95\% CI: 0.84 - 0.95). This is a 5.7\% absolute increase (a 6.7\% relative improvement) in AUC performance ($p$-value $=0.000$). 
  Furthermore, the AP score improves from 0.62 (95\% CI: 0.51 - 0.75) to 0.78 (95\% CI: 0.66 - 0.89), a 16.0\% absolute increase (25.9\% relative gain, $p$-value $=0.000$). 
  In addition to improving AUC performance, our proposed method exhibits a substantial AP gain, indicating that the population-aware module not only enhances overall separability but also delivers markedly higher precision (fewer false positives) across all recall levels.

  \zsubsection{Varying neighborhood size $k$}
 We further examine how the hypergraph neighborhood size $k$, thus the population-awareness context size, affects anomaly discrimination. We compare the AE under three configurations: 1) The baseline or plain AE, which is equivalent to $k=0$, where there is no population-level context, 2) A \emph{small-k} configuration with $k = \lceil 0.35 \space B\rceil $, which is 35\% of the batch-size, and half the context capacity of the proposed solution, 3). A \emph{large-k} configuration, representing the proposed solution so far, where 70\% of the batch size is utilized when estimating the $k$-nearest neighbors of a node ($k = \lceil 0.70 \space B \rceil $). We set the batch size $B = 16$, which translates to the values $k = {0, 6, 12}$, respectively. 
 
 \Cref{fig:ad-scores-dist} and ~\Cref{fig:ad-roc-pr-acc-all} visualize the distribution plots and AUC-ROC/AP performance with each of the three configurations as a series in the plots, while~\Cref{tbl:acc-perf-ci} details the statistical test results. 
 A visual examination of the distribution plots shows increasing separability between healthy and pathological scores as $k$ grows. However, the KS statistics for $k=0$ and \emph{small-$k$} are equivalent (KS = 0.7 in each case). 
 Similarly, AUC scores for $k=0$ and \emph{small-$k$} are equivalent at 0.84 ($p$-value = 0.791), with performance gains being statistically different under \emph{large-$k$} only. 
 As for the AP score, however, this progressive gain in performance as $k$ increases is also observed between $k=0$ and \emph{small-$k$}, where the latter yields a 2.7\% relative improvement that is also statistically significant ($p$-value=0.000), though numerically not substantial (1\% absolute change compared to 16\% absolute improvement under \emph{large-$k$}). 
 This appears to reiterate that the proposed relational regularization particularly improves precision.

 Additionally, the results suggest that the relationship between $k$ and performance gains is not linear; reducing the context size by half (from 70\% of batch to 35\% of batch) does not exhibit a proportionate difference in AUC and AP. More importantly, it points to a graceful degradation, where \emph{small-$k$} performs similarly to the baseline $k=0$, when there is insufficient neighborhood context to leverage.  
 The relational regularization also requires sufficient neighborhood context, in our case $k = 0.7B$, to substantially improve discrimination performance, presenting an area for further investigation concerning optimal $k$.

\begin{figure}[htbp]  
    \centering  
    \begin{subfigure}[b]{0.32\textwidth}
        \centering 
        \includegraphics[width=\linewidth]{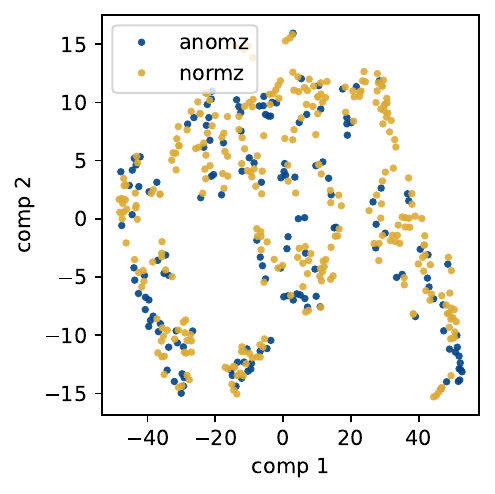}
\subcaption{Plain AE ($k=0$)}\label{fig:zc-scatter-vanilla}\end{subfigure}
\begin{subfigure}[b]{0.32\textwidth}
        \centering 
        \includegraphics[width=\linewidth]{02-zc-cluster-qlty--scatter-obspop\_35-umap-valid}
\subcaption{AE+Small-$k$}\label{fig:zc-scatter-obspop-35}\end{subfigure}
\begin{subfigure}[b]{0.32\textwidth}
        \centering 
        \includegraphics[width=\linewidth]{02-zc-cluster-qlty--scatter-obspop\_70-umap-valid}
\subcaption{AE+Large-$k$}\label{fig:zc-scatter-obspop-70}\end{subfigure}
\hfill

    \caption{Latent code space} 
     
    \label{fig:latent-scatter} 
      
    \end{figure}

 \zsubsection{Latent space quality}
 To assess how the proposed relational context restructures the latent space, we evaluate clustering quality across neighborhood sizes $k = \{0, 6, 12\}$ (baseline, \emph{small-$k$}, \emph{large-$k$}, respectively).
 ~\Cref{fig:latent-scatter} visualizes the latent codes via UMAP, showing better isolation of pathological samples under \emph{large-$k$}.

 \begin{table}[htbp!]
\centering
\caption{Cluster quality of resulting latent space}
\label{tbl:latent-cluster-qlty}
\small
\begin{tabular}{lp{1.3in}p{1.3in}p{1.3in}c}
\toprule
Clustering metric & Plain AE ($k=0$) & AE+Small-$k$ & AE+Large-$k$ & Trend \\
\midrule
Adjusted mutual info score & 0.00 \scriptsize{$\pm 0.03 $} & 0.01 \scriptsize{$\pm 0.03 $} & 0.04 \scriptsize{$\pm 0.06 $} & \minitrend{0.0}{0.004321373782642578}{0.01703333929878037} \\
Adjusted rand score & -0.01 \scriptsize{$\pm 0.02 $} & -0.01 \scriptsize{$\pm 0.02 $} & 0.01 \scriptsize{$\pm 0.03 $} & \minitrend{-0.004364805402450088}{-0.004364805402450088}{0.004321373782642578} \\
Calinski harabasz score & 1.00 \scriptsize{$\pm 1.50 $} & 1.83 \scriptsize{$\pm 2.29 $} & 4.19 \scriptsize{$\pm 2.66 $} & \minitrend{0.3010299956639812}{0.45178643552429026}{0.7151673578484579} \\
Davies bouldin score & 13.73 \scriptsize{$\pm 39.85 $} & 20.78 \scriptsize{$\pm 41.80 $} & 10.38 \scriptsize{$\pm 11.45 $} & \minitrend{1.1682027468426308}{1.338057875419756}{1.0561422620590524} \\
Homogeneity score & 0.15 \scriptsize{$\pm 0.09 $} & 0.09 \scriptsize{$\pm 0.06 $} & 0.13 \scriptsize{$\pm 0.07 $} & \minitrend{0.06069784035361165}{0.037426497940623665}{0.05307844348341968} \\
Silhouette score & -0.32 \scriptsize{$\pm 0.28 $} & -0.25 \scriptsize{$\pm 0.15 $} & -0.04 \scriptsize{$\pm 0.03 $} & \minitrend{-0.16749108729376372}{-0.12493873660829995}{-0.017728766960431602} \\
\bottomrule
\end{tabular}
\end{table}

 Quantitatively,~\Cref{tbl:latent-cluster-qlty} reports mean results and 95\% confidence intervals of clustering metrics. While the scores point to existing overlaps, four of six metrics improve monotonically with $k$. For the other two metrics (without a monotonic trend), the \emph{large-$k$} configuration still achieves the best performance. For instance, it has the best Davies-Bouldin score (lower is better) with the narrowest confidence interval, indicating both improved cluster separation and stability.
 All in all, integrating the population-aware relational context with sufficient $k$ yields a more discriminative and structured latent representation.

\zsubsection{Downstream application}
\Cref{fig:ad-roc-pr-acc-all} shows the classification performance using both the anomaly scores and the latent codes from each of the three $k$ sizes.~\Cref{tbl:acc-perf-ci} and~\Cref{tbl:dstream-perf-ci} detail the mean results and associated confidence intervals and statistical tests.

\renewcommand{\arraystretch}{0.95} 
\begin{table}[htbp!]
\centering
\caption{ROC, PR, and downstream accuracy performance @ thresholded anomaly score}
\label{tbl:acc-perf-ci}
\small
\begin{tabular}{p{0.6in}cccccc}
\toprule
Model & AUC & AP & Accuracy & F1 & Precision & Recall \\
\midrule
Plain AE ($k=0$) & \makecell[t]{
 $\mathbf{ 0.84 }$ \\ \text{\scriptsize ($0.77-0.91$) } \\ [2.0ex]} & \makecell[t]{
 $\mathbf{ 0.62 }$ \\ \text{\scriptsize ($0.51-0.75$) } \\ [2.0ex]} & \makecell[t]{
 $\mathbf{ 0.84 }$ \\ \text{\scriptsize ($0.77-0.89$) } \\ [2.0ex]} & \makecell[t]{
 $\mathbf{ 0.79 }$ \\ \text{\scriptsize ($0.73-0.86$) } \\ [2.0ex]} & \makecell[t]{
 $\mathbf{ 0.70 }$ \\ \text{\scriptsize ($0.61-0.79$) } \\ [2.0ex]} & \makecell[t]{
 $\mathbf{ 0.92 }$ \\ \text{\scriptsize ($0.84-0.98$) } \\ [2.0ex]} \\
AE+Small-$k$ & \makecell[t]{
 $\mathbf{ 0.84 }$ \raisebox{0.5ex}{\text{\tiny ns }} \\ \text{\scriptsize ($0.77-0.90$) } \\ \text{\scriptsize $\blacktriangledown \mathbf{ 0.0\% }$, } \text{\scriptsize $-0.0\%$ \textit{rel.} } \\ [2.0ex] } & \makecell[t]{
 $\mathbf{ 0.63 }$ \raisebox{0.5ex}{\text{\tiny *** }} \\ \text{\scriptsize ($0.52-0.76$) } \\ \text{\scriptsize $\blacktriangle \mathbf{ 1.7\% }$, } \text{\scriptsize $2.7\%$ \textit{rel.} } \\ [2.0ex] } & \makecell[t]{
 $\mathbf{ 0.84 }$ \raisebox{0.5ex}{\text{\tiny *** }} \\ \text{\scriptsize ($0.78-0.89$) } \\ \text{\scriptsize $\blacktriangle \mathbf{ 0.4\% }$, } \text{\scriptsize $0.5\%$ \textit{rel.} } \\ [2.0ex] } & \makecell[t]{
 $\mathbf{ 0.79 }$ \raisebox{0.5ex}{\text{\tiny *** }} \\ \text{\scriptsize ($0.72-0.86$) } \\ \text{\scriptsize $\blacktriangledown \mathbf{ 0.3\% }$, } \text{\scriptsize $-0.4\%$ \textit{rel.} } \\ [2.0ex] } & \makecell[t]{
 $\mathbf{ 0.71 }$ \raisebox{0.5ex}{\text{\tiny *** }} \\ \text{\scriptsize ($0.61-0.81$) } \\ \text{\scriptsize $\blacktriangle \mathbf{ 1.2\% }$, } \text{\scriptsize $1.7\%$ \textit{rel.} } \\ [2.0ex] } & \makecell[t]{
 $\mathbf{ 0.89 }$ \raisebox{0.5ex}{\text{\tiny *** }} \\ \text{\scriptsize ($0.80-0.98$) } \\ \text{\scriptsize $\blacktriangledown \mathbf{ 2.9\% }$, } \text{\scriptsize $-3.2\%$ \textit{rel.} } \\ [2.0ex] } \\
AE+Large-$k$ & \makecell[t]{
 $\mathbf{ 0.90 }$ \raisebox{0.5ex}{\text{\tiny *** }} \\ \text{\scriptsize ($0.84-0.95$) } \\ \text{\scriptsize $\blacktriangle \mathbf{ 5.7\% }$, } \text{\scriptsize $6.7\%$ \textit{rel.} } \\ [2.0ex] } & \makecell[t]{
 $\mathbf{ 0.78 }$ \raisebox{0.5ex}{\text{\tiny *** }} \\ \text{\scriptsize ($0.66-0.89$) } \\ \text{\scriptsize $\blacktriangle \mathbf{ 16.0\% }$, } \text{\scriptsize $25.9\%$ \textit{rel.} } \\ [2.0ex] } & \makecell[t]{
 $\mathbf{ 0.88 }$ \raisebox{0.5ex}{\text{\tiny *** }} \\ \text{\scriptsize ($0.83-0.93$) } \\ \text{\scriptsize $\blacktriangle \mathbf{ 4.2\% }$, } \text{\scriptsize $5.0\%$ \textit{rel.} } \\ [2.0ex] } & \makecell[t]{
 $\mathbf{ 0.83 }$ \raisebox{0.5ex}{\text{\tiny *** }} \\ \text{\scriptsize ($0.76-0.90$) } \\ \text{\scriptsize $\blacktriangle \mathbf{ 4.0\% }$, } \text{\scriptsize $5.1\%$ \textit{rel.} } \\ [2.0ex] } & \makecell[t]{
 $\mathbf{ 0.77 }$ \raisebox{0.5ex}{\text{\tiny *** }} \\ \text{\scriptsize ($0.68-0.87$) } \\ \text{\scriptsize $\blacktriangle \mathbf{ 7.8\% }$, } \text{\scriptsize $11.2\%$ \textit{rel.} } \\ [2.0ex] } & \makecell[t]{
 $\mathbf{ 0.91 }$ \raisebox{0.5ex}{\text{\tiny *** }} \\ \text{\scriptsize ($0.82-0.98$) } \\ \text{\scriptsize $\blacktriangledown \mathbf{ 1.7\% }$, } \text{\scriptsize $-1.8\%$ \textit{rel.} } \\ [2.0ex] } \\
 & \minitrend{0.2653598617874564}{0.2653053983342506}{0.27848219327902635} & \minitrend{0.20891882525225935}{0.21345095480461504}{0.2498787070107902} & \minitrend{0.264285268052691}{0.26520718251471725}{0.2740583063290131} & \minitrend{0.2535782442412787}{0.25278543132258685}{0.26319399294393253} & \minitrend{0.22963096404254846}{0.2326638482013813}{0.24910388675761821} & \minitrend{0.28387101239486606}{0.27724842703987446}{0.28012229565758817} \\
\bottomrule
\end{tabular}
\end{table}

\begin{table}[htbp!]
\centering
\caption{ROC, PR, and downstream accuracy performance @ regress on latent code}
\label{tbl:dstream-perf-ci}
\small
\begin{tabular}{p{0.5in}cccccc}
\toprule
Model & AUC & AP & Accuracy & F1 & Precision & Recall \\
\midrule
Plain AE ($k=0$) & \makecell[t]{
 $\mathbf{ 0.61 }$ \\ \text{\scriptsize ($0.49-0.72$) } \\ [2.0ex]} & \makecell[t]{
 $\mathbf{ 0.48 }$ \\ \text{\scriptsize ($0.36-0.62$) } \\ [2.0ex]} & \makecell[t]{
 $\mathbf{ 0.67 }$ \\ \text{\scriptsize ($0.61-0.72$) } \\ [2.0ex]} & \makecell[t]{
 $\mathbf{ 0.22 }$ \\ \text{\scriptsize ($0.06-0.38$) } \\ [2.0ex]} & \makecell[t]{
 $\mathbf{ 0.55 }$ \\ \text{\scriptsize ($0.25-0.95$) } \\ [2.0ex]} & \makecell[t]{
 $\mathbf{ 0.14 }$ \\ \text{\scriptsize ($0.03-0.26$) } \\ [2.0ex]} \\
AE+Small-$k$ & \makecell[t]{
 $\mathbf{ 0.80 }$ \raisebox{0.5ex}{\text{\tiny *** }} \\ \text{\scriptsize ($0.72-0.89$) } \\ \text{\scriptsize $\blacktriangle \mathbf{ 19.2\% }$, } \text{\scriptsize $31.3\%$ \textit{rel.} } \\ [2.0ex] } & \makecell[t]{
 $\mathbf{ 0.66 }$ \raisebox{0.5ex}{\text{\tiny *** }} \\ \text{\scriptsize ($0.51-0.81$) } \\ \text{\scriptsize $\blacktriangle \mathbf{ 18.0\% }$, } \text{\scriptsize $37.3\%$ \textit{rel.} } \\ [2.0ex] } & \makecell[t]{
 $\mathbf{ 0.78 }$ \raisebox{0.5ex}{\text{\tiny *** }} \\ \text{\scriptsize ($0.70-0.85$) } \\ \text{\scriptsize $\blacktriangle \mathbf{ 10.8\% }$, } \text{\scriptsize $16.1\%$ \textit{rel.} } \\ [2.0ex] } & \makecell[t]{
 $\mathbf{ 0.65 }$ \raisebox{0.5ex}{\text{\tiny *** }} \\ \text{\scriptsize ($0.52-0.76$) } \\ \text{\scriptsize $\blacktriangle \mathbf{ 42.9\% }$, } \text{\scriptsize $193.8\%$ \textit{rel.} } \\ [2.0ex] } & \makecell[t]{
 $\mathbf{ 0.70 }$ \raisebox{0.5ex}{\text{\tiny *** }} \\ \text{\scriptsize ($0.56-0.87$) } \\ \text{\scriptsize $\blacktriangle \mathbf{ 15.2\% }$, } \text{\scriptsize $27.6\%$ \textit{rel.} } \\ [2.0ex] } & \makecell[t]{
 $\mathbf{ 0.61 }$ \raisebox{0.5ex}{\text{\tiny *** }} \\ \text{\scriptsize ($0.43-0.75$) } \\ \text{\scriptsize $\blacktriangle \mathbf{ 47.0\% }$, } \text{\scriptsize $329.5\%$ \textit{rel.} } \\ [2.0ex] } \\
AE+Large-$k$ & \makecell[t]{
 $\mathbf{ 0.85 }$ \raisebox{0.5ex}{\text{\tiny *** }} \\ \text{\scriptsize ($0.76-0.92$) } \\ \text{\scriptsize $\blacktriangle \mathbf{ 23.3\% }$, } \text{\scriptsize $38.1\%$ \textit{rel.} } \\ [2.0ex] } & \makecell[t]{
 $\mathbf{ 0.75 }$ \raisebox{0.5ex}{\text{\tiny *** }} \\ \text{\scriptsize ($0.62-0.88$) } \\ \text{\scriptsize $\blacktriangle \mathbf{ 27.2\% }$, } \text{\scriptsize $56.5\%$ \textit{rel.} } \\ [2.0ex] } & \makecell[t]{
 $\mathbf{ 0.81 }$ \raisebox{0.5ex}{\text{\tiny *** }} \\ \text{\scriptsize ($0.74-0.88$) } \\ \text{\scriptsize $\blacktriangle \mathbf{ 14.1\% }$, } \text{\scriptsize $21.0\%$ \textit{rel.} } \\ [2.0ex] } & \makecell[t]{
 $\mathbf{ 0.72 }$ \raisebox{0.5ex}{\text{\tiny *** }} \\ \text{\scriptsize ($0.63-0.83$) } \\ \text{\scriptsize $\blacktriangle \mathbf{ 50.2\% }$, } \text{\scriptsize $226.8\%$ \textit{rel.} } \\ [2.0ex] } & \makecell[t]{
 $\mathbf{ 0.72 }$ \raisebox{0.5ex}{\text{\tiny *** }} \\ \text{\scriptsize ($0.60-0.86$) } \\ \text{\scriptsize $\blacktriangle \mathbf{ 17.2\% }$, } \text{\scriptsize $31.4\%$ \textit{rel.} } \\ [2.0ex] } & \makecell[t]{
 $\mathbf{ 0.73 }$ \raisebox{0.5ex}{\text{\tiny *** }} \\ \text{\scriptsize ($0.58-0.87$) } \\ \text{\scriptsize $\blacktriangle \mathbf{ 58.7\% }$, } \text{\scriptsize $411.1\%$ \textit{rel.} } \\ [2.0ex] } \\
 & \minitrend{0.20758973500929137}{0.2564460428979387}{0.2662509269671907} & \minitrend{0.17062703930055464}{0.22034376197583813}{0.243798803910943} & \minitrend{0.22309495507909793}{0.2502863346601039}{0.2583456518310584} & \minitrend{0.0868040413181144}{0.21750775835905375}{0.23631978272390164} & \minitrend{0.1902137644849236}{0.23083871510613785}{0.236032029701341} & \minitrend{0.05793502965676526}{0.20760520838722493}{0.23787832692562053} \\
\bottomrule
\end{tabular}
\end{table}

\renewcommand{\arraystretch}{1.0} 

The threshold-based classifier (using Youden's $J$ statistic) exhibits relatively larger jumps in F1 and precision scores at larger $k$ (F1 (0.79→0.83, +5.1\%), and precision (0.70→0.77, +11.6\%)), while maintaining high recall (0.91) that is numerically equivalent to the baseline recall score of 0.92. 

Logistic classification on the latent codes yields inconsistent performance for the baseline AE (AUC = 0.61 [CI: 0.49 - 0.72],  F1 = 0.22 [CI: 0.06 - 0.38]), indicating poor linear separability. 
In contrast, under \emph{large-$k$} configuration, the proposed solution produces a structured latent space that is readily linearly separable, attaining an F1-score of 0.72 and AUC 0.85. Incremental gains are observed as $k$ increases from \emph{small-$k$} to \emph{large-$k$}. 
This demonstrates that the proposed relational regularization not only improves anomaly scores but also yields a latent representation that is more amenable to direct linear classification.

\begin{figure}[htbp]
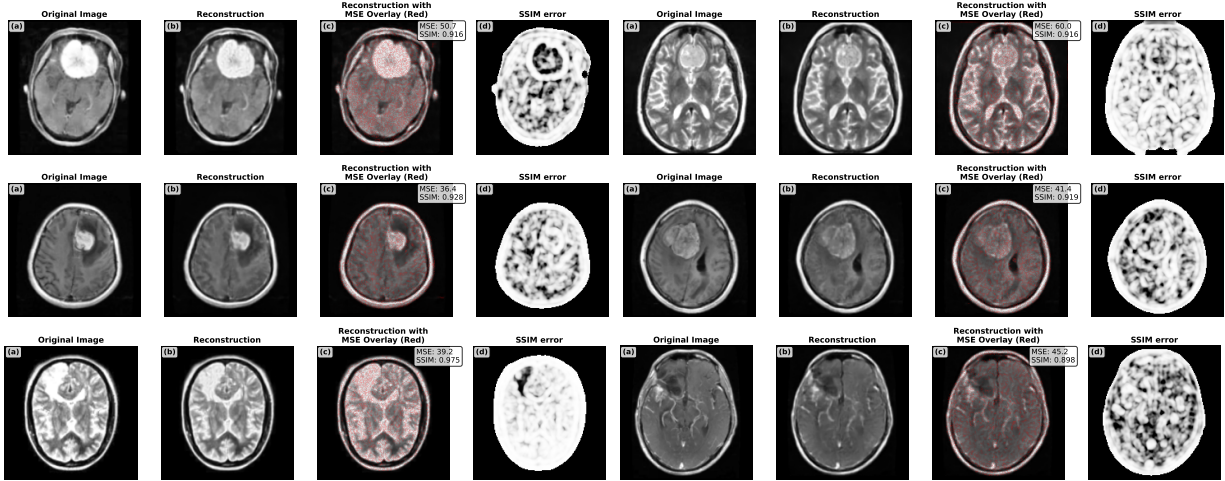
  
    \centering  
    \begin{subfigure}[b]{0.49\textwidth}
        \centering 
        \includegraphics[width=\linewidth]{obspop\_70-anomz-267}
\end{subfigure}
\begin{subfigure}[b]{0.49\textwidth}
        \centering 
        \includegraphics[width=\linewidth]{obspop\_70-anomz-247}
\end{subfigure}
\hfill
\begin{subfigure}[b]{0.49\textwidth}
        \centering 
        \includegraphics[width=\linewidth]{obspop\_70-anomz-272}
\end{subfigure}
\begin{subfigure}[b]{0.49\textwidth}
        \centering 
        \includegraphics[width=\linewidth]{obspop\_70-anomz-255}
\end{subfigure}
\hfill
\begin{subfigure}[b]{0.49\textwidth}
        \centering 
        \includegraphics[width=\linewidth]{obspop\_70-anomz-167}
\end{subfigure}
\begin{subfigure}[b]{0.49\textwidth}
        \centering 
        \includegraphics[width=\linewidth]{obspop\_70-anomz-213}
\end{subfigure}
\hfill

    \caption{Reconstruction of anomalous regions} 
     
    \label{fig:ae-recon-anomz} 
      
    \end{figure}

\paragraph{Reconstruction of anomalous regions}
Qualitative inspection of reconstructed images, as exemplified in \Cref{fig:ae-recon-anomz}, demonstrates failure to completely suppress reconstruction of anomalous regions. Despite the gains in distributional separation and precision in anomaly detection, the proposed method does not entirely prevent the decoder from generating anatomically coherent outputs for pathological inputs.

\section{Discussion}
\label{sec:ddiscuss}

This work introduces a relational latent regularization module that enhances unsupervised anomaly detection by integrating a population-level context into the bottleneck layer of an auto-encoder. 
Our evaluation validates this module, demonstrating the feasibility of the proposed in-batch integration and the improved ability to separate pathological anomalies from healthy anatomical variation.

The primary contribution of the proposed approach is a marked improvement in precision (reduction in false positive rates). While overall discriminability (as measured by AUC) improved by 5.7\% absolute change, AP increased by 16.0\% (25.9\% relative improvement). This disproportionate gain suggests that the module particularly enhances specificity in addition to the general capacity to separate the distributions. The downstream threshold-based classifier reiterates this pattern, showing a 7.8\% absolute improvement in precision with minimal impact on recall.

Our sensitivity analysis reveals that performance gains depend on a sufficiently large relational neighborhood. A small neighborhood ($k = \lceil 0.35B \rceil$) yields an AUC score that is statistically indistinguishable from the baseline ($k$ = 0). Significant improvements emerge only with a larger context ($k = \lceil 0.70 B\rceil$), where the AP score increases by 25.9\% relative to the baseline performance. 
This indicates that capturing a robust sense of normalcy requires integrating information from a substantial subset of the batch. 
Furthermore, the repeated pattern of monotonic improvement in various metrics with increasing $k$ suggests an optimizable parameter, with a potential sweet spot for maximum gain. 

Visualization and clustering metrics show that the proposed method transforms the latent representation, producing a latent space with better separation and cluster coherence when sufficient neighborhood context (under \emph{large-$k$}) is incorporated. Most tellingly, a simple logistic regression classifier achieves an F1-score of 0.72 on the latent code produced by the proposed method, compared to 0.22 for the plain AE, demonstrating an induced representation that is not only better for reconstruction but also readily linearly separable for downstream diagnostic tasks. 

Finally, while the proposed module demonstrates enhanced anomaly detection performance, it remains limited in its ability to suppress visually perceptible reconstructions of anomalous tissue.


\paragraph{Limitations.} 
The optimal neighborhood size $k$ likely depends on dataset size and heterogeneity, and, therefore, a theoretical or adaptive method for setting $k$ would be valuable. Relatedly, $k$ is also a function of the batch size, which additionally introduces computational overhead during graph estimation at training time. This, too, could benefit from more granular investigation. 
Furthermore, while we demonstrate performance on a heterogeneous neuroimaging dataset, validation across more anatomies and imaging modalities is necessary.

\section{Conclusion}
\label{sec:dconc} 
 
 The proposed relational-based population-level context operates on the principle that normalcy in biological samples is inherently heterogeneous. 
 By enforcing consistency within a local neighborhood of healthy samples, the module learns a more robust manifold of normal anatomy, leading to improvements in precision, latent space quality, and downstream utility. 
 The improvements in anomaly detection particularly increase the reduction of false positives, attaining 0.90 (95\% CI: 0.84 - 0.95) AUC and 0.78 (95\% CI: 0.66 - 0.89) AP scores, which is a 5.7\% and 16.0\% absolute improvement on the baseline performance, respectively. 
 Integration of the module into the bottleneck layer of a CNN auto-encoder, with dynamic estimation of an in-batch-based relational neighborhood, additionally offers a flexible method for integration.

\section*{Acknowledgements}

{
	\bibliographystyle{unsrtnat} 
	\bibliography{\dpaperbib}
}


\end{document}